\documentclass[aps,pra,twocolumn,superscriptaddress,longbibliography]{revtex4-2}

\pdfoutput=1

\usepackage{amsmath}
\usepackage{graphicx}
\usepackage{color}
\usepackage{amsfonts}
\usepackage{amssymb}
\usepackage{braket}
\usepackage{times}
\usepackage{lineno}
\usepackage{siunitx}
\usepackage{lineno}
\usepackage{hyperref}
\usepackage{url}
\usepackage{xcolor}
\hypersetup{colorlinks = true, citebordercolor={blue}, linkcolor={blue}, citecolor={blue}, urlcolor={blue}}

\begin{document}

\title{\huge Quantum super-resolution microscopy by photon statistics and structured light}

\author{F. Picariello}
\email{fabio.picariello@polito.it}
\affiliation{\textit{Politecnico di Torino, Department of Electronics and Telecommunications, Italy}}
\affiliation{\textit{Istituto Nazionale di Ricerca Metrologica, Italy}}

\author{E. Losero}
\affiliation{\textit{Istituto Nazionale di Ricerca Metrologica, Italy}}

\author{S. Ditalia Tchernij} 
\affiliation{\textit{ Physics Dept., University of Torino, Italy}}
\affiliation{\textit{Istituto Nazionale di Fisica Nucleare sez. Torino, Italy}}

\author{P. Boucher}
\affiliation{\textit{Quantonation, Paris, France}}

\author{M.Genovese}
\affiliation{\textit{Istituto Nazionale di Ricerca Metrologica, Italy}}
\affiliation{\textit{ Physics Dept., University of Torino, Italy}}

\author{I. Ruo-Berchera}
\affiliation{\textit{Istituto Nazionale di Ricerca Metrologica, Italy}}

\author{I. P. Degiovanni}
\affiliation{\textit{Istituto Nazionale di Ricerca Metrologica, Italy}}
\affiliation{\textit{ Physics Dept., University of Torino, Italy}}

\begin{abstract}
We present an advanced quantum super-resolution imaging technique based on photon statistics measurement and its accurate modeling. Our reconstruction algorithm adapts to any kind of non-Poissonian emitters, outperforming the corresponding classical SOFI method. It offers sub-diffraction resolution improvement that scales with the $\sqrt{j}$, where $j$ is the highest order central moments of the photocounts. More remarkably, in combination with structured illumination a linear improvement with j can be reached.
Through simulations and experiments, we prove our method's clear superiority over traditional SOFI, especially in low excitation light conditions, providing a promising avenue for non-invasive super-resolution microscopy of delicate samples.

\end{abstract}

\maketitle

\section{Introduction}
Improving spatial resolution is critical in current biological and medical research \cite{Sahl2017}. Within the principle of optical diffraction, Ernst Abbe initially identified the optical resolution limits as imposed by the numerical aperture of the objective and the wavelength of light \cite{abbe1873beitrage}. For more than a century it was considered an impassable limit. However, in the early 1990s, novel approaches broke beyond Abbe's limit by circumventing some of its underlying assumptions (explicit or implicit), among them the linear and static response of the sample to incident light, the far-field intensity detection and uniform illumination. In particular, by exploiting the non-linear or randomly selective response of point-like fluorescence markers, pioneering techniques such as stimulated emission depletion (STED) microscopy and single-molecule localization microscopy (SMLM) have substantially advanced the field and demonstrated exceptional resolution enhancement \cite{hell1994breaking,lelek2021single,betzig2006imaging,xu2021recent}. 

However, when using photosensitive samples these methods can be of limited use. In particular, biologists often demand live cell imaging technologies that deliver high resolution while minimizing photodamage led to the creation of super-resolution techniques, which posed some questions about the reliability of these methods for bio-compatible purposes \cite{marx2013super,laissue2017assessing,waldchen2015light,mauranyapin2017evanescent}.

 A bio-compatible super-resolution approach, that does not require high illumination levels, is the super-resolution optical fluctuation microscopy (SOFI) \cite{dertinger2009fast,cevoli2021design,pawlowska2021embracing,classen2018analysis}, which exploits the natural or induced random fluctuation of fluorophores' brightness. However, as we will discuss later, SOFI is based on a semi-classical model of light that only contemplates classical super-Poisson emitters, does not take into account quantum fluctuation, and leads to a strongly limited resolution enhancement in low light scenarios. 
 Especially in low light regime, the quantization of light in photons rapidly emerges and 
 only a full quantum description provides the route to reach the optimal performance in optical imaging \cite{genovese2016real}, often leveraging on peculiar quantum features. For instance, sub-shot noise imaging and sensing with enhanced signal-to-noise ratio can be achieved by exploiting entangled light beams \cite{brida2010experimental,brida2011experimental,Berchera_2019,Samantaray_2017,Ortolano_2021a,Ortolano_2023b,Ortolano_2023a,Zhang_2024}. The genuine quantum anti-bunching properties of single photon emitters have been used to achieve super-resolution by measuring higher order correlation function of the photo-counts both in wide-field  \cite{schwartz2013superresolution} and in confocal settings \cite{gatto2014beating}. Those last methods can be seen as the equivalent of SOFI but for sub-Poissonian emitters.

This paper aims to introduce a full quantum method for super-resolution named quantum super-resolution imaging by photon statistics (QSIPS) that is based on a rigorous full quantum model describing the photon emission and detection. Our model yields to a generalized SOFI approach that works optimally at any intensity level and with any non-Poissonian emitter, including single photon emitters and any kind of blinking (photoswitching) fluorophores. Here we present the model together with simulations, and experimental results, demonstrating the superiority of our approach over traditional SOFI methodologies published in the literature. The QSIPS super-resolved image is actually the incoherent sum of the power $j$ of each emitter's point spread function (PSF), where $j$ is the highest order central moments evaluated at each point of the image plane. Thus, assuming a Gaussian PSF the application of QSIPS provides an enhancement of the resolution of a factor $\sqrt{j}$, corresponding to the effective narrowing of the function $\text{(PSF)}^{j}$. 

Furthermore, we present the integration of our QSIPS method, with Structured Illumination Microscopy (SIM) \cite{gustafsson2000surpassing}. In SIM the sample is illuminated with a set of non-uniform patterns, and the final super-resolved image is reconstructed by the combination of all the single pattern images. SIM doubles the maximum frequency information retrieved by the optical system. The integration of non-Poissonian super-resolution methods with SIM has been demonstrated theoretically to provide a much more favorable $j+\sqrt{j}$ super-resolution scaling with the correlation order \cite{classen2017superresolution} and the principle has been applied experimentally both for wide field SOFI \cite{descloux2021experimental} and for quantum image scanning microscopy \cite{tenne2019super}. 

Here, we demonstrate that even in combination with structured light our QSIPS method outperforms the classical implementation with SOFI, especially in scenarios characterized by low light illumination.

\section{Super-resolution model}
Consider a system of $N_c$ emitters with mutually incoherent and statistically independent emissions. Let $P_\alpha(m)$ denote the probability of emitting $m$ photons from the $\alpha$-th source. Each photon possesses a certain probability of being detected at position $\mathbf{r}$ in the detector plane, denoted as $\eta_\alpha(\mathbf{r}) = \rho_\alpha PSF_\alpha(\mathbf{r})$, where $\rho_\alpha$ accounts for all the optical losses in the setup, including the non-unit efficiency of the detector. Here, $PSF_\alpha(\mathbf{r})$ represents the imaged system point-spread-function related to the $\alpha$-th emitter.

The effect of the position-dependent loss setup, $\eta_\alpha(\mathbf{r})$, alters the detected photon distribution at the image plane. In particular, the detected distribution for the $\alpha$-th emitter, denoted as $\mathcal{P}_\alpha(n,\mathbf{r})$ can be expressed according to a Binomial statistical model as:
\begin{align}
\mathcal{P}_\alpha(n,\mathbf{r}) = \sum_{m=n}^{\infty} P_\alpha(m) \binom{m}{n}\left[\eta_\alpha(\mathbf{r})\right]^n[1-\eta_\alpha(\mathbf{r})]^{(m-n)}, \label{Pk}
\end{align}
while considering the contribution of all the emitters at position $\mathbf{r}$ we have:
\begin{align}
    \mathcal{P}(N,\mathbf{r}) =  &\left(\sum_{n_1=0,\dots n_{Nc}}^{\infty} \mathcal{P}_1(n_1,\mathbf{r})\dots \mathcal{P}_{Nc}(n_{Nc},\mathbf{r})\right) \times \label{PN} \\ \nonumber & \times \delta_{N,\sum\limits_\alpha n_\alpha},
\end{align}
that is, for each position $\mathbf{r}$, the convolution of all the emitters contributions.

We can identify a statistical quantity known as cumulant. In particular, the $j$-th cumulant is obtained performing the $j$-th derivative of the cumulant generating function $\mathcal{K}(t,\mathbf{r})$ evaluated at $t=0$:
\begin{align}
\mathcal{K}_\eta(t,\mathbf{r}) &= \log\left(\sum_{N=0}^{\infty}e^{tN}\mathcal{P}(N,\mathbf{r})\right),\\
k_\eta^{(j)}(\mathbf{r}) &= \frac{d^j\mathcal{K}}{dt^j}(t,\mathbf{r})|_{_{t=0}}.
\label{cumulantdetected}
\end{align}
$k_\eta^{(j)}(\mathbf{r})$ refers to the $j$-th cumulant of the overall detected photon distribution evaluated in the position $\mathbf{r}$ of the detector plane that, for independent emitters can be written as the sum of each emitter cumulants: $k_{\alpha,\eta}^{(j)}(\mathbf{r})$:
\begin{equation}
    k_{\eta}^{(j)}(\mathbf{r})=\sum_{\alpha=1}^{N_c}k_{\alpha,\eta}^{(j)}(\mathbf{r}).
\end{equation}
The classical SOFI technique relies on assessing the auto-cumulants of the detected photon distribution \cite{dertinger2009fast}. Therefore, the $j$-th order super-resolved signal is evaluated as:
\begin{equation}
    SOFI^{(j)}(\mathbf{r}) := k_\eta^{(j)}(\mathbf{r})=\sum_{\alpha=1}^{N_c}k_{\alpha,\eta}^{(j)}(\mathbf{r}).
\end{equation}
This classical approach however does not properly take into account the quantum noise deriving from the discrete nature of the photons. To properly understand what is the origin of the limit of the SOFI technique we write the second and third-order super-resolved map evaluated with classical SOFI, explicating the PSF contributions $\eta_\alpha(\mathbf{r})$ to the signal:
\begin{align}
    SOFI^{(2)}(\mathbf{r}) &=\sum_{\alpha=1}^{N_c} \left[\left[\eta_\alpha(\mathbf{r})\right]^2\left(z^{(2)}_\alpha-z_\alpha^{(1)}\right) + \eta_\alpha(\mathbf{r})z^{(1)}_\alpha\right], \label{sofi2}\\
    SOFI^{(3)}(\mathbf{r})  &=\sum_{\alpha=1}^{N_c} \left[\left[\eta_\alpha(\mathbf{r})\right]^3\left(z^{(3)}_\alpha-3z_\alpha^{(2)}+2z_\alpha^{(1)}\right)\right. + \label{sofi3}\\ \nonumber & \quad \quad \quad \left.+ 3\left[\eta_\alpha(\mathbf{r})\right]^2\left(z_\alpha^{(2)}-z_\alpha^{(1)}\right)+ \right.\\ \nonumber & \left. \quad \quad \quad -2\eta_\alpha(\mathbf{r}) z^{(1)}_\alpha\right], \nonumber
\end{align}
with the terms $z_\alpha^{(j)}$ being the cumulant of a generic order $j$ of the emitted photon distribution $P_\alpha(m)$, evaluated starting from the cumulant generating function $Z_\alpha(t)$:
\begin{align}
    Z_\alpha(t) &= \log\left(\sum_{m=0}^{\infty}e^{tm}P_\alpha(m) \right),\\
    z_\alpha^{(j)}&=\frac{d^jZ_\alpha}{dt^j}(t)|_{_{t=0}}.
\end{align}
Note that, in  Eq. (\ref{sofi2}),  the only presence of the terms proportional to $[\eta_\alpha(\mathbf{r})]^2$ would correspond to a shrink of a factor $\sqrt{2}$ of each PSF and would lead to a corresponding super-resolved image. However, the other terms, linear with $\eta_\alpha(\mathbf{r})$, are what limits the maximum resolution enhancement. These linear terms originate from the randomness of a loss process when taking into account the discreteness of photons. The same consideration holds for the achievement of third-order super-resolution in Eq. (\ref{sofi3}), which is limited by the presence of terms in $\eta_\alpha(\mathbf{r})$ and in $[\eta_\alpha(\mathbf{r})]^2$.

For the second-order signal in Eq. (\ref{sofi2}), we can see that second-order SOFI fails when the linear terms in $\eta_\alpha(\mathbf{r})$ dominates. Considering that the first order cumulant is the mean value of the emitted photon distribution, $z_\alpha^{(1)}=\langle m_{\alpha} \rangle$, and the second order cumulant is the variance, $z_\alpha^{(2)}=\langle \Delta^{2} m_{\alpha} \rangle$, the condition becomes:
\begin{equation}
\eta_\alpha(\mathbf{r})\left(F^{(e)}_\alpha-1\right) \leq 1, \label{SOFIapproxQSIPS1}
\end{equation}
where we have introduced the Fano factor of the emitted photon distribution relative to the $\alpha$-th emitter as: $F^{(e)}_\alpha:= \langle \Delta^{2} m_{\alpha}\rangle/\langle m_{\alpha} \rangle $.
Finally,  using the general relation between the Fano factor before and after an optical loss $\eta$,  $F^{(d)}=\eta( F^{(e)}-1)+ 1$,  we can rearrange Eq. (\ref{SOFIapproxQSIPS1}) in terms of the Fano factor of the detected photon distribution as:
\begin{equation}
    F^{(d)}_\alpha(\mathbf{r})\leq 2.\label{Fanofactor_detected}
\end{equation}
This condition can be used as a rule of thumb to understand when the SOFI technique fails to attain the maximum resolution enhancement due to contributions of lower order power in the shrinking of the PSF. 

The regime dictated by Eq.s (\ref{SOFIapproxQSIPS1}) and (\ref{Fanofactor_detected}) is a characteristic either of high losses system (i.e. extremely low values of $\eta_\alpha(\mathbf{r})$) or of sub-Poissonian emitters, i.e. with $0\leq F^{(e)}<1$. Particularly, we can see this by considering that an ideal single photon emitter possesses a probability density function such that all the cumulants are identically null, except the first order one (i.e the mean value of the distribution). Subsequently, the second and third-order SOFI signal become:
\begin{align}
    SOFI^{(2)}(\mathbf{r}) &=\sum_{\alpha=1}^{N_c}\eta_\alpha(\mathbf{r})z_\alpha^{(1)}\left[1-\eta_\alpha(\mathbf{r})\right],\label{SOFI2SPS}\\
    SOFI^{(3)}(\mathbf{r})  &=\sum_{\alpha=1}^{N_c} 2\eta_\alpha(\mathbf{r})z_\alpha^{(1)}\left[\left[\eta_\alpha(\mathbf{r})\right]^2-\frac{3}{2}\eta_\alpha(\mathbf{r})-1\right].\label{SOFI3SPS}
\end{align}
Since $\eta_\alpha(\mathbf{r}) < 1$, all SOFI signals evaluated for a system of single photon emitters do not provide super-resolved maps due to the dominance of the first order of the PSF respect to higher ones.

We here propose a new algorithm for super-resolution, the QSIPS, that overcomes the above described SOFI limitations, efficiently eliminating the lower order contributions in the shrinking of the PSF from the super-resolved signal and therefore extending its usage to sub-Poissonian photon sources.

In particular, the complete super-resolved signal of order $j$, $\text{QSIPS}^{(j)}$, can be obtained by performing a linear combination of cumulants of the detected photon distribution, as: 
\begin{align}
    QSIPS^{(j)}(\mathbf{r}) &:= \sum_{i=1}^{j}\beta_{i,j}k_\eta^{(i)}(\mathbf{r}) \label{QSIPS} \\ \nonumber &=\sum_{\alpha=1}^{N_c}\left[\eta_{\alpha}(\mathbf{r})\right]^j\left(\sum_{i = 0}^{j}\beta_{i,j}z^{(i)}_{\alpha}\right),
\end{align}
where we demonstrate that the coefficients $\beta_{i,j}$ coincides with the Stirling number of the first type \cite{broder1984r}:
\begin{equation}
\beta_{i,j} = \frac{1}{j!}\frac{d^j}{dx^j}\prod_{i=0}^{j-1}(x-i)|_{x=0}=S_I(i,j).\label{eqcoefficient}
\end{equation}
Eq. (\ref{QSIPS}) shows how to properly combine cumulants up to the $j$-th order to derive a signal proportional uniquely to $[\eta_\alpha(\mathbf{r})]^j$. 
This super-resolution method applies to all kinds of non-Poissonian photon sources and it is completely independent on the optical losses, therefore ensuring the maximum super-resolution enhancement in all levels of illumination.

Another way to rewrite Eq. (\ref{QSIPS}) more compactly is to exploit the following generating function, labeled Sgurzant generating function:
\begin{equation}
\mathcal{S}_\eta(\rho,\mathbf{r})=\log\left(\sum_{N=0}^{\infty}\rho^N\mathcal{P}(N,\mathbf{r})\right).\label{Sgenerating}
\end{equation}
The peculiarity of this generating function is that its moment of order $j$, evaluated performing the $j$-th derivative of Eq. (\ref{Sgenerating}) at $\rho=1$, retrieves the exact linear combination of cumulants of the detected photon distribution from order $1$ to $j$, reported in Eq. (\ref{QSIPS}) with coefficients $\beta_{i,j}$.
Therefore, the super-resolved relation of order $j$ reported in Eq. (\ref{QSIPS}) can be rewritten as:
\begin{equation}
    QSIPS^{(j)}(\mathbf{r}) = \frac{d^j\mathcal{S}_\eta}{d\rho^j}(\rho,\mathbf{r})|_{_{\rho=1}}.\label{SgurzantQSIPS}
\end{equation}
The proof of Eq.s (\ref{QSIPS})-(\ref{SgurzantQSIPS}), with the requirement of non-Poissonianity of the sources, is reported in the Supplement 1.

In our experiment, we exploited, as a specific example of non-Poissonian photon sources, quantum dots (QDs) nano-particles that exhibit blinking behavior.
Even though QDs can be treated as single photon emitters, the presence of trap states can lead the photon source to a temporary "off" state, which emits non-radiatively \cite{efros2016origin}. This blinking characteristic is what gives rise to super-Poissonian photon statistics.

To ensure proper interpretation of the experimental results we carry on simulations where we have considered QDs with super-Poissoniann behavior as the ones in our experiment and with sub-Poissonian as ideal single photon emitters.

We considered the QDs as independent emitters with uncorrelated emissions that perform $M_\alpha$ excitation cycles in a single exposure time. The blinking nature of the emitters is considered by associating to each excitation cycle a certain probability $b_\alpha$, noted as blinking probability, in which, upon a single excitation cycle, no photon is emitted. The probability density function related to the emission of a generic emitter $\alpha$ is modeled as:
\begin{equation}
    P_\alpha(m) = b_\alpha\delta_{m,0}+(1-b_\alpha)\delta_{m,M_\alpha}.\label{simulPDF}
\end{equation}
The emitted photon distribution Fano factor can be evaluated as $F^{(e)}_\alpha = M_\alpha b_\alpha$, with the condition $M_\alpha b_\alpha>1$ ensuring super-Poissonian statistics, while the condition reported in Eq. (\ref{SOFIapproxQSIPS1}) becomes:
\begin{equation}
\eta_\alpha(\mathbf{r})\left(M_\alpha b_\alpha-1\right)\leq 1.
\end{equation}

\section{Results and discussion}
A simulation was conducted in MATLAB on three closely spaced, identical emitters each with the probability density function $P_\alpha(m)$ reported in Eq. (\ref{simulPDF}). The detection scheme consists of a matrix of ideal, noise-free photon number resolving detectors, in an optical system with unitary magnification, each pixel characterized by a probability of detecting $N$ photons $\mathcal{P}(N,\mathbf{r})$ according to Eq. (\ref{PN}). Individual detector images were temporally averaged to generate a classical intensity map $\langle N \rangle$, while the super-resolution images were obtained by assessing auto-cumulant operation separately to each pixel using both QSIPS and SOFI techniques.

\begin{figure}[!hb]
\centering
\includegraphics[width=1\linewidth]{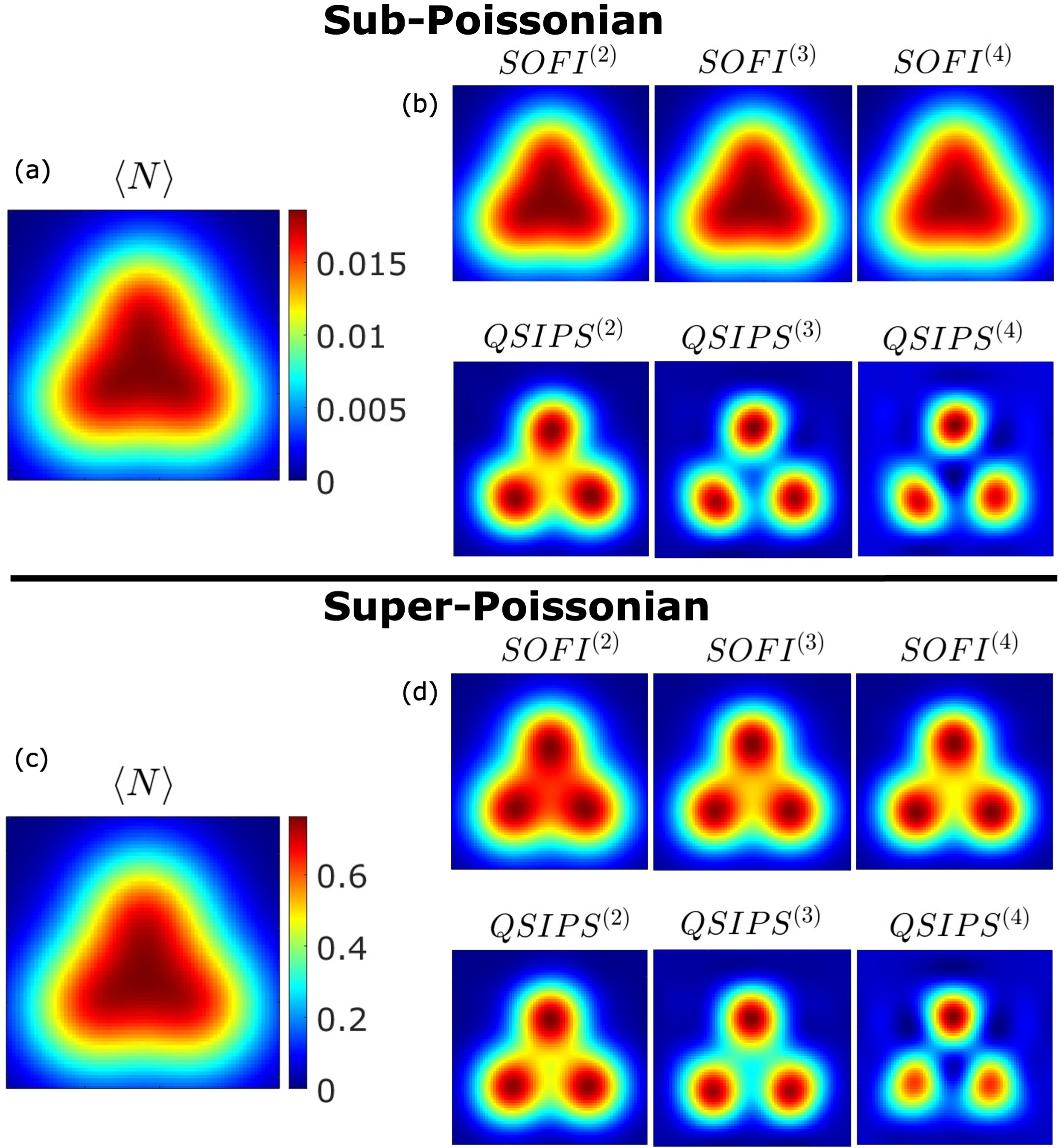}
\caption{Simulation of three identical non-Poissonian emitters in a noiseless scenario with standard deviation of the PSF ($\sigma$) of $1.2$ and spaced $1.15\times \sigma$ in pixel units. (a) The mean number of detected photons from ideal non-blinking single photon emitters ($b_\alpha=0$, $M_\alpha = 1$ for any $\alpha$). (b) SOFI and QSIPS super-resolved images up to the order four for single-photon emitters evaluated by $3 \times 10^6$ independent frames. (c) Mean number of detected photons for blinking super-Poissonian emitters ($b_\alpha=0.7$, $M_\alpha = 70$ for any $\alpha$). (d)  SOFI and QSIPS super-resolved images up to the order four for blinking emitters evaluated by $1 \times 10^5$ independent frames. All simulations are performed by setting an optical loss 
$1-\rho_\alpha=0.75$, which does not include the further effect of PSF's spatial spreading.  Fourier interpolation and Gaussian filtering have been applied to improve image quality and reduce noise.}
\label{QSIPSsimul}
\end{figure}

Fig \ref{QSIPSsimul} shows two distinct cases: the top panel presents both super-resolution methods, up to the fourth order, employing ideal not blinking single photon emitters (i.e. sub-Poissonian photon sources), while the bottom panel refers to the presence of blinking characteristic that gives rise to super-Poissonian behavior. Particularly, we can notice that in the sub-Poissonian case, the SOFI method fails in providing any resolution enhancement, as suggested by Eq.s (\ref{SOFI2SPS}) and (\ref{SOFI3SPS}), while in the super-Poissonian one, the increase in resolution is only partially achieved. 
Nevertheless, our QSIPS methods demonstrate its ability to enhance the spatial resolution regardless the photon statistics considered.

\begin{figure}[!hb]
\centering
\includegraphics[width=1\linewidth]{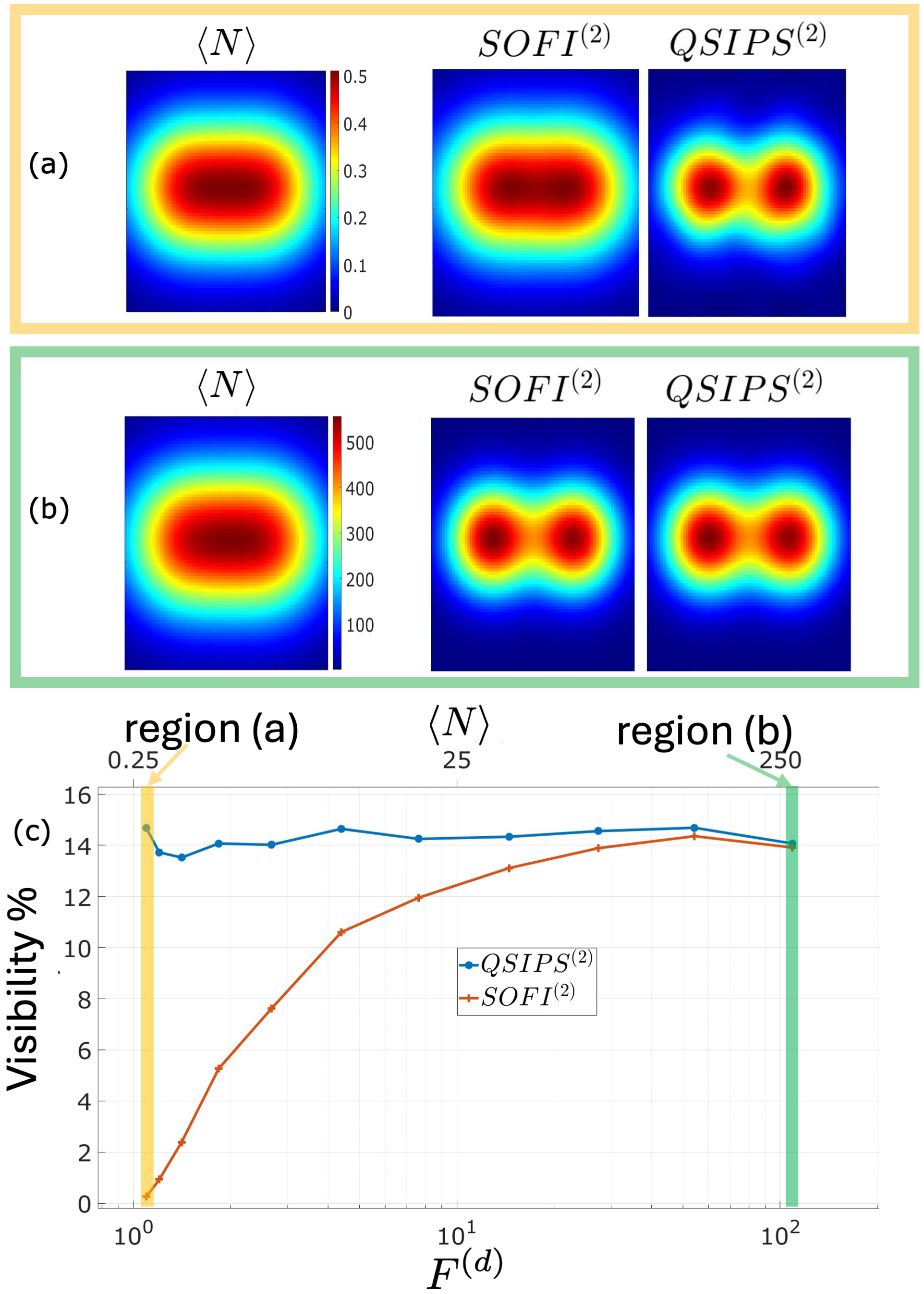}
\caption{Visibility comparison of two identical blinking single photon emitters (simulated) with standard deviation of the PSF ($\sigma$) of $1.55$ and spaced $1\times \sigma$ in pixel units. (a) From left to right: mean intensity map, second-order SOFI, and second-order QSIPS maps in the low photon number regime, obtained by setting $M_\alpha = 60$ and $b_\alpha = 0.3$ for both emitters. (b) it shows the same quantities, albeit in a strong light regime, with $M_\alpha = 60\times10^3$, $b_\alpha = 0.3$. (c) Visibility of the dip between the emitters in function of the Fano factor (bottom axis) and the mean number of detected photons (top axis) obtained by changing the number of excitations per simulated frame from $M_\alpha=60$ to $M_\alpha=60\times10^3$. Both the bottom and the top axes refer to the average values of the respective quantities evaluated in the region covered by the PSF of the emitters. The parameters chosen in the examples of panels (a) and (b) correspond to highlighted yellow and green bars in panel (c), respectively. All simulations are performed by setting an optical loss 
$1-\rho_\alpha=0.85$, which does not include the further effect of PSF's spatial spreading and a statistical sample of $10^5$ frames. The images displayed were Fourier interpolated to reduce pixelation effects.}
\label{QSIPSvsSOFIvisibility}
\end{figure}

In Fig. \ref{QSIPSvsSOFIvisibility}, we report the simulation study of the super-resolution visibility in the case of super-Poissonian emitters. The visibility is accessed in terms of the relative deepness of the valley between two identical blinking emitters obtained in the second-order super-resolved images via SOFI and QSIPS methods (see the examples reported in Fig. \ref{QSIPSvsSOFIvisibility} (a)-(b)).

Fig. \ref{QSIPSvsSOFIvisibility} (c) shows that QSIPS maintains constant visibility at all levels of illumination and for all values of the detected Fano factor. In contrast, the visibility gained with the traditional SOFI method grows with the number of detected photons and the corresponding Fano factor, according to the discussion of Eq. (\ref{Fanofactor_detected}), finally reaching the QSIPS's performance for $F^{(d)}\gg 1$. 

\begin{figure*}[!ht]
\centering
\includegraphics[width=0.9\linewidth]{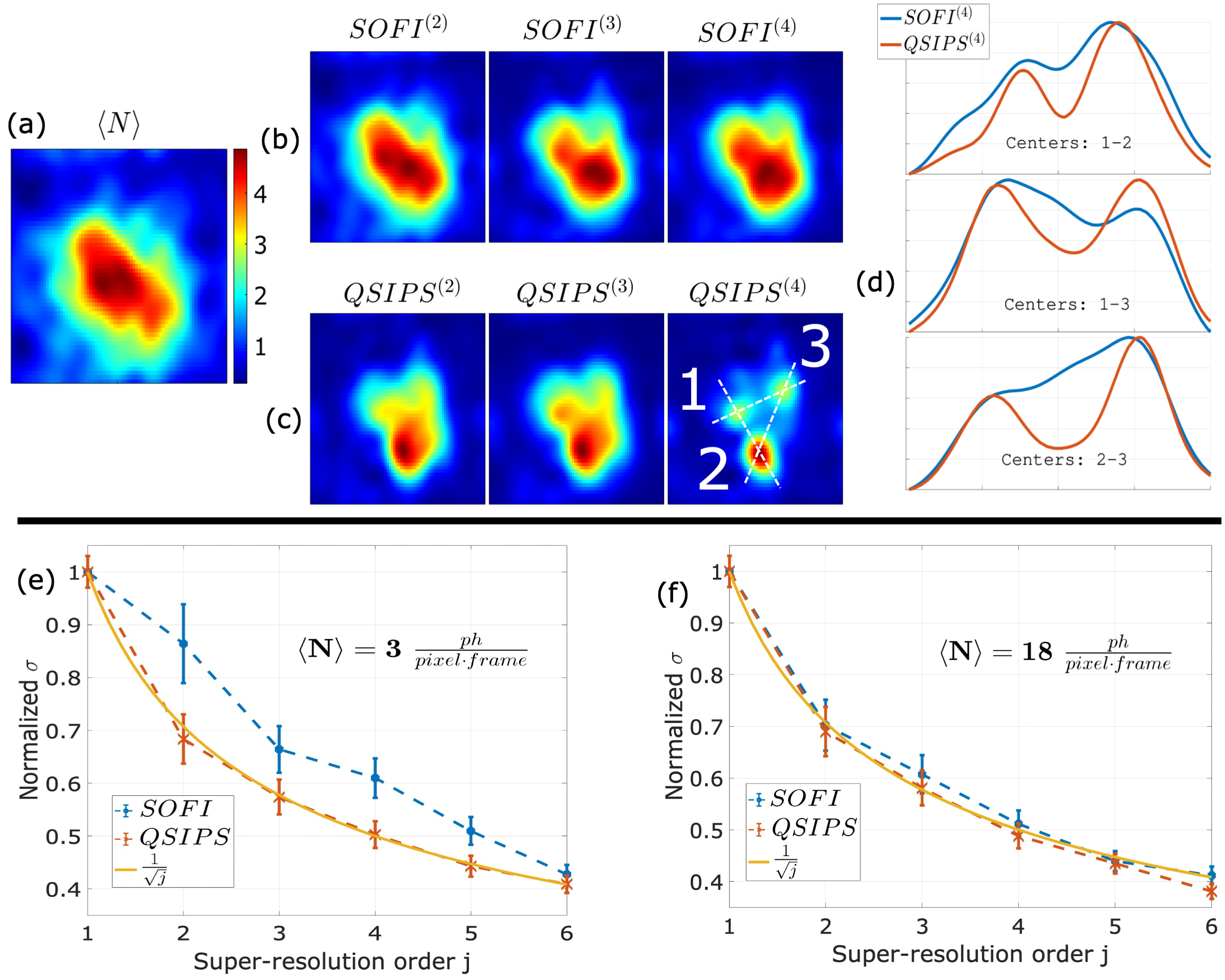}
\caption{Experimental comparison between SOFI and QSIPS methods. (a) Mean intensity map of a QDs cluster. (b) - (c): first four super-resolution orders' maps obtained using the SOFI and QSIPS approaches, respectively. (d) Normalized fourth-order super-resolution signals in a 1D plot along intersecting axes from the three centers indicated in panel (c). An exposure duration of \SI{20}{\milli\second} and a laser power of \SI{1.4}{\milli\watt} were used to capture $53 \times 10^3$ frames. 
Fourier interpolation and Gaussian filtering were used to improve image quality and minimize noise. 
(e)-(f). The $y$-axis shows the (normalized) standard deviation of the effective PSF evaluated for an isolated emitter depending on super-resolution orders (x-axis) under two light conditions: $\langle N \rangle = 3$ ph/pixel/frame at $50$ ms exposure, and $\langle N \rangle = 18$ ph/pixel/frame at $200$ ms exposure, respectively. The SOFI (blue dashed line) and QSIPS (orange dashed line) data are represented. The theoretical behavior is shown as a continuous yellow line.}\label{QSIPS_vs_SOFI_exp}
\end{figure*}

An experimental demonstration of the superiority of QSIPS with respect to SOFI has been performed employing the setup described in Section \ref{Methods}. In particular, Fig.s \ref{QSIPS_vs_SOFI_exp} (a)-(d) display the intensity map along with the super-resolved images of a small cluster of QDs, and a 1D plot to showcase the resolution enhancement. In particular, the fourth-order QSIPS reconstruction shows the inner structure of the cluster that cannot be unveiled by the standard SOFI. 

To quantify the real super-resolution improvement, we investigated the narrowing of the effective (super-resolved) PSF for an isolated emitter in function of the super-resolution order $j$. The analysis is based on a 2D Gaussian fit of the effective PSFs applied to each super-resolution order. A normalization of the size of each super-resolved PSFs with the one of the intensity's PSF is applied. The result is illustrated in Fig. \ref{QSIPS_vs_SOFI_exp} (e) and (f), which refer to two different exposure times. The experimental data for SOFI (orange stars) and QSIPS (blue dots) are compared with the theoretical expected scaling of $1/\sqrt{j}$ (yellow curve). We can see that in the high illumination regime of  Fig. \ref{QSIPS_vs_SOFI_exp} (f), both approaches coincide with the theoretical prediction; however, with a limited number of detected photons as in Fig. \ref{QSIPS_vs_SOFI_exp} (e), the actual resolution boost offered by SOFI is inconsistent across multiple super-resolution orders.

Although this study does not provide experimental results utilizing sub-Poissonian light sources, a version of the QSIPS method, applicable exclusively to single-photon emitters, has been proposed and experimentally demonstrated \cite{gatto2014beating}. Supplement 1 illustrates that the mathematical model reported in \cite{gatto2014beating} is a specific instance of our generalized QSIPS method.

\subsection{Integration with Structured Light}

\begin{figure*}[!ht]
\centering
\includegraphics[width=0.9\linewidth]{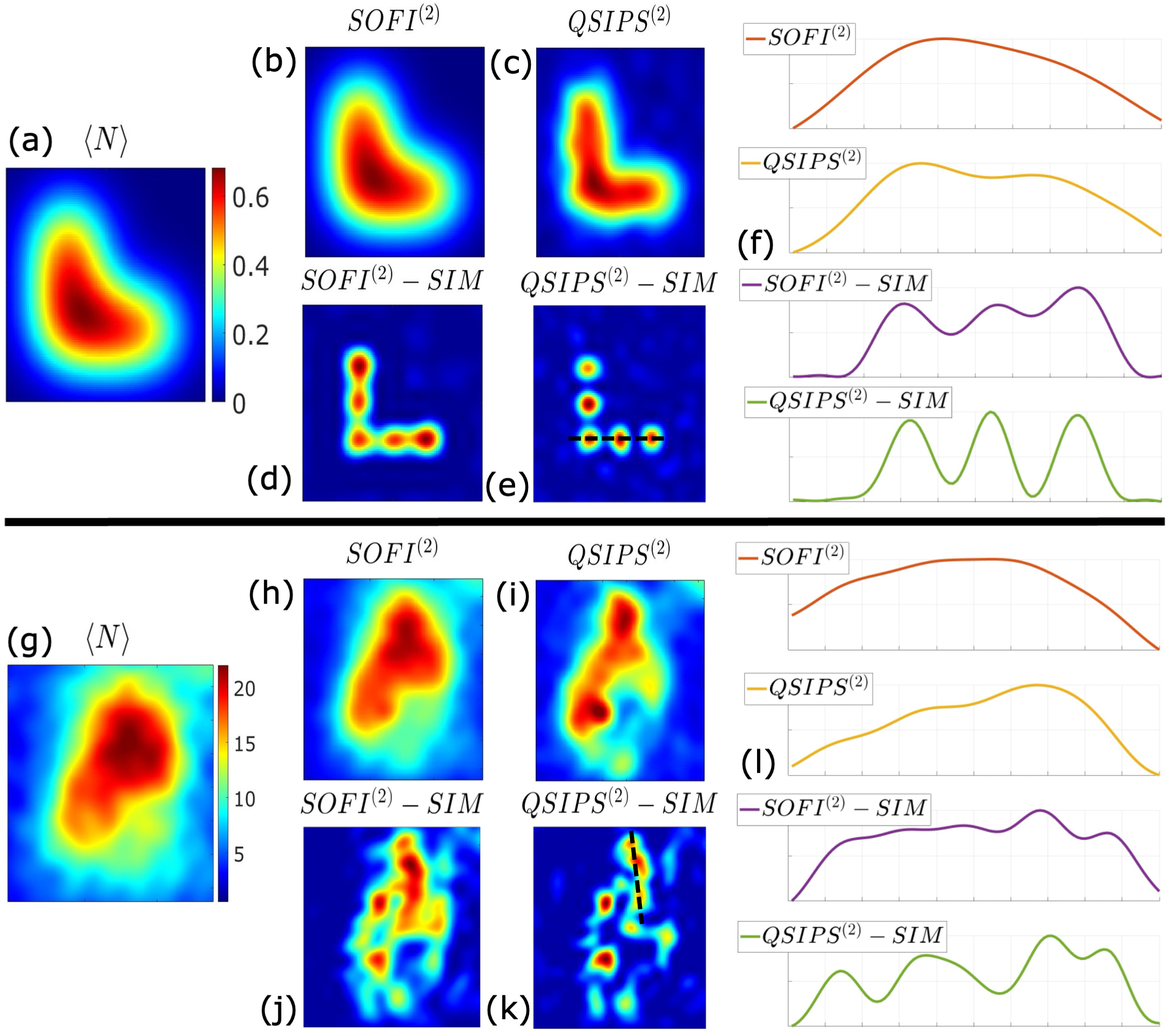}
\caption{Second-order SOFI and QSIPS algorithms integrated with SIM. Panels (a)–(f) show the simulation results for an L-shaped configuration of identical blinking single-photon emitters using a noisy detection system. Panel (a) shows the mean intensity map. Panels (b) and (c) show second-order super-resolved pictures from the $\text{SOFI}^{(2)}$ and $\text{QSIPS}^{(2)}$ approaches, respectively. Panels (d) and (e) demonstrate the integration of the SOFI and QSIPS methods with SIM, labeled $\text{SOFI}^{(2)}-\text{SIM}$ and $\text{QSIPS}^{(2)}-\text{SIM}$, respectively. Panel (f) displays a normalized 1D plot along the dashed line in panel (e) for all super-resolution approaches. The simulations assume $90\%$ optical losses (excluding the PSF effect), a blinking probability $b_\alpha = 0.1$, $M_\alpha = 100$ excitation cycles per frame for all the emitters, RMS readout noise of $0.23$, and $5 \times 10^3$ independent frames per value of $\theta$ and $\phi$. Panels (g)–(l) refer to the same quantities of the top panels, albeit for an experimental realization performed on a cluster of colloidal QDs. The experimental setup utilized a \SI{200}{\milli\second} exposure duration, \SI{500}{\micro\watt} laser power, and $5 \times 10^2$ frames per value of $\theta$ and $\phi$. To increase quality, all pictures underwent Fourier interpolation and Gaussian filtering.}
\label{SIM_integration}
\end{figure*}

The fundamental concept behind SIM is to employ an illumination pattern such as:
\begin{equation}
    I_{\theta,\phi}(\mathbf{r}) = \frac{I_0}{2}\left[1-\cos(2\pi\left(\mathbf{p}_\theta\cdot\mathbf{r}\right)+\phi)\right],\label{Intensity_structured}
\end{equation}
with $\mathbf{r}$ denoting the spatial position on the object plane and $\mathbf{p}_\theta$ referring to the illumination frequency vector.
Combining different acquisitions performed with different values of $\theta$ and $\phi$ by shifting the Fourier components of $\pm |\mathbf{p}_\theta|$ allows to cover twice the Fourier space normally achievable with standard illumination, therefore doubling the resolution enhancement \cite{lal2016structured}. 

In this work, we focus on integrating second-order super-resolved maps evaluated through non-Poissonian photon statistics with structured light. 

The greatest resolution improvement is obtained when the magnitude of $|\mathbf{p}_{\theta}|$ equals the maximum spatial frequency that the system can broadcast, $k_{\text{Abbe}}$. This frequency is related to the width, $\sigma$, of the system's optical PSF as $k_{\text{Abbe}} = 0.42/\sigma$ \cite{zhang2007gaussian}. The next simulations will use $|\mathbf{p}_{\theta}|=k_\text{Abbe}$.

In the following, we present the results of simulations carried on in conditions analogous to the one of Fig. \ref{QSIPSsimul}, but for the sake of completeness, we also added the contribution of the readout noise to each detection pixel modeled as a Gaussian distribution convoluted with the detected photon statistics. The readout noise represents the main noise contribution in low light level situations when employing classical wide-field linear sensors \cite{bigas2006review}. To add the influence of structured illumination, the emission probability of $k$ photons by each $\alpha$-th emitter becomes:
\begin{equation}
    \pi_{\alpha,\theta,\phi}(k) = \sum_{m=0}^{\infty} P_{\alpha}(m) B\left(k \mid m, I_{\theta,\phi}(x_{\alpha}, y_{\alpha})\right),
\end{equation}
where $I_{\theta,\phi}(x, y)$ derives from Eq. (\ref{Intensity_structured}) by setting  $I_0=1$, \\ $\mathbf{p}_\theta = |\mathbf{p}_\theta|(\cos(\theta),\sin(\theta))$ and $\mathbf{r}=(x,y)$. Thus, structured light operates as a binomial filter with a transmittance probability of $I_{\theta,\phi}(x_{\alpha}, y_{\alpha})$.

The probability of observing $n$ photons at position $\mathbf{r}$ is analogous to Eq. (\ref{Pk}) where the probability $\pi_{\alpha,\theta,\phi}(k)$ plays the role of $P_{\alpha}(m)$. In particular, $\pi_{\alpha,\theta,\phi}(k)$ undergoes a binomial process driven by the combination of PSF and optical losses $\eta_{\alpha}(x, y)$.

Using the detected photon distributions obtained in this way, a series of twenty simulated datasets is created, considering four values of $\theta$ and five values of $\phi$. For each value of $\theta$ and $\phi$, second-order super-resolved images are generated using both QSIPS and SOFI methods. Each non-Poissonian super-resolved map encompasses spatial frequency information within a radius of $\sqrt{2}k_{\text{Abbe}}$. By accurately shifting each frequency component in the Fourier space it is possible to extend the coverage in the reciprocal space up to a factor $2+\sqrt{2}$ with respect to the region achievable with conventional illumination. 
\cite{classen2017superresolution}. 

The values of $\theta$ and $\phi$ employed in these simulations, as well as in the experimental realization, were:
\begin{align}
    \theta &= \left\{0, \frac{\pi}{4}, \frac{\pi}{2},\frac{3\pi}{4}\right\} + \frac{\pi}{8}, \\ \phi &= \left\{0, \frac{2\pi}{5}, \frac{4\pi}{5}, \frac{6\pi}{5}, \frac{8\pi}{5}\right\} + \frac{\pi}{8}.
\end{align}
First, for each of the twenty combinations of \(\theta\) and \(\phi\), the mean intensity map and the second-order super-resolved SOFI and QSIPS maps are evaluated. The corresponding averages of the twenty maps are reported in  Fig. \ref{SIM_integration} (a), (b), and (c).

Then, the images at the various orientations and phases, are Fourier-transformed, Wiener-filtered, and shifted in the Fourier space according to the SIM algorithm \cite{lal2016structured}, using the assessed frequency vector $\mathbf{p}_\theta$. Finally, the frequency components are summed and inverse Fourier-transformed.
The final super-resolved images labeled as $\text{SOFI}^{(2)}-\text{SIM}$ and $\text{QSIPS}^{(2)}-\text{SIM}$, are shown in Fig. \ref{SIM_integration} (d) and (e). The 1D plot along the dashed black line in (e) is reported in panel (f). The simulation clearly shows a resolution enhancement of QSIPS with respect to SOFI maintained when combined with structured illumination.

The setup described in Section \ref{Methods} was used to experimentally integrate SIM with the QSIPS approach in low-light scenarios. Preliminarily, a high-density sample of colloidal QDs was used to assess the illumination frequency vectors $\mathbf{p}_\theta$  for all orientations. Following that, the same sample used for the analysis in Fig. \ref{QSIPS_vs_SOFI_exp} was employed, but focusing on a different cluster whose intensity image is shown in Fig. \ref{SIM_integration} (g). 

As for the simulation scenario, the $\text{SOFI}^{(2)}$ and $\text{QSIPS}^{(2)}$ are evaluated for each of the twenty structured illumination patterns, and their average is reported in Fig. \ref{SIM_integration} (h)-(i). Afterward, they are combined in Fourier space and transformed back into the real space leading to the super-resolved images in Fig. \ref{SIM_integration} (j)-(k). These results demonstrate also experimentally the advantage of QSIPS-SIM in comparison with the SOFI-SIM.

\begin{figure}[!ht]
\centering
\includegraphics[width=1\linewidth]{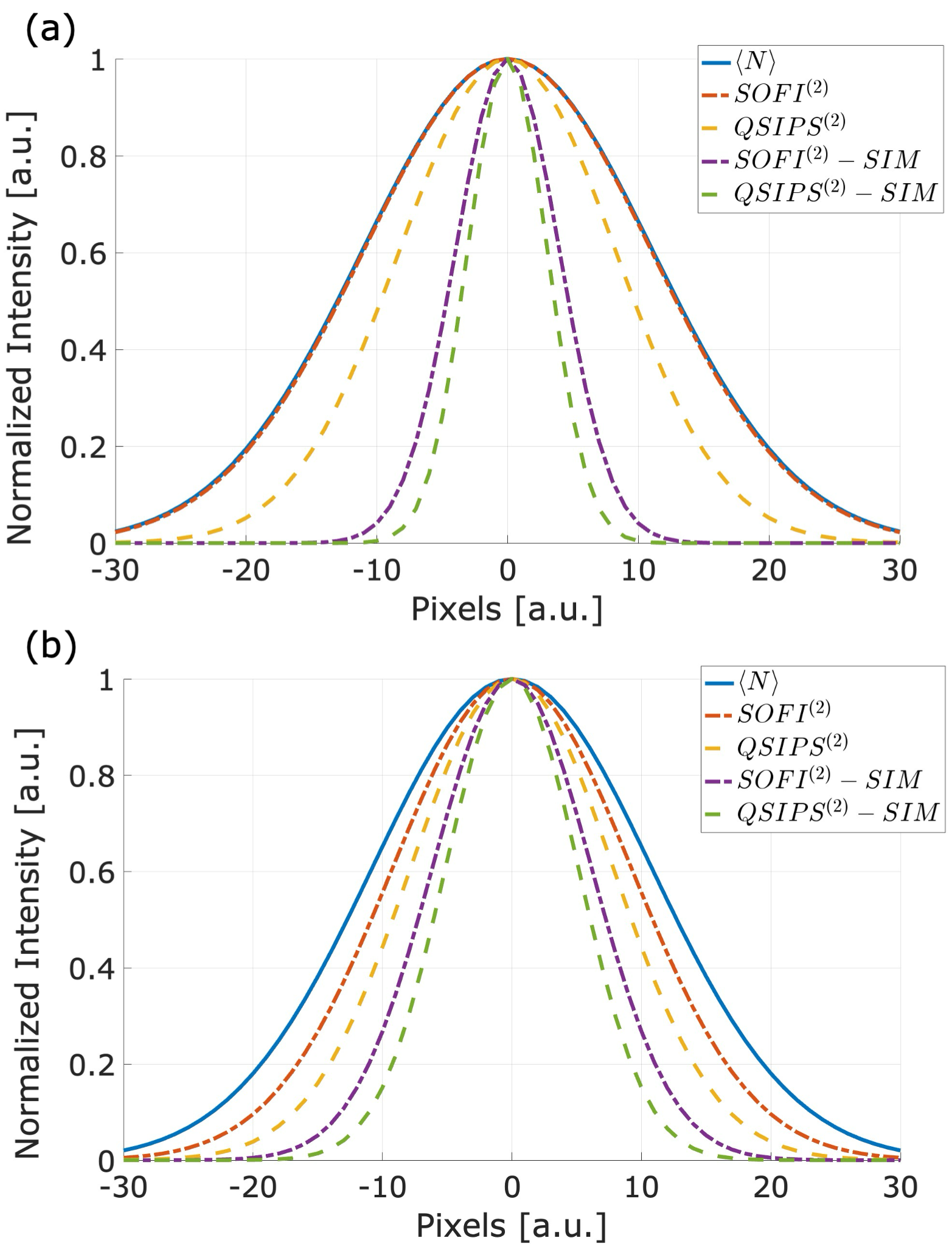}
\caption{1D Gaussian fit of the effective PSF of a single emitter. Panel (a) depicts a simulated emitter in a noisy detection system, with blinking probability $b_\alpha=0.1$, $M_\alpha=10^2$ excitation cycles per frame, and Gaussian readout noise $RMS = 0.23$. Panel (b) shows the same analysis performed experimentally on a single-emitter with the same settings as per Fig. \ref{SIM_integration} (g)-(l).}
\label{SE_experimental}
\end{figure}

To assess the actual super-resolution provided by the integration with structured light,
we focused on an isolated emitter and performed a two-dimensional Gaussian fit of the effective PSFs across multiple super-resolution techniques. Particularly, we evaluated the super-resolution enhancement by performing the ratio between the PSF's standard deviation of the intensity image and the one obtained for all the different super-resolution methods. Fig. \ref{SE_experimental} shows the 1D Gaussian fits for both simulation (panel (a))  and experimental realization (panel (b)).
The super-resolution enhancement for both simulated and experimental results, as their theoretical values, are reported in Table \ref{resolution_factors}. 

\begin{table*}[!ht]
\centering
\caption{\bf Super-resolution enhancement}
\begin{tabular}{ccccc}
\hline
  & $SOFI^{(2)}$ & $QSIPS^{(2)}$ & $SOFI^{(2)}-SIM$ & $QSIPS^{(2)}-SIM$  \\
\hline
\textbf{Simul} $\left(F^{(d)} = 1.03 \pm 0.02\right)$ & $1.01\pm0.02$ & $1.40\pm0.02$ &$2.63\pm0.02$ &$3.42\pm0.02$ \\
\textbf{Exp} $\left( F^{(d)} = 1.8 \pm 0.1\right)$ & $1.2\pm0.1$ & $1.4\pm0.1$ & $1.8\pm0.1$ & $2.3\pm0.1$\\
\textbf{Th} & $1.4$ & $1.4$ & $3.4$& $3.4$ \\
\hline
\end{tabular}
  \label{resolution_factors}
\end{table*}

The super-resolution enhancement factor for $\text{SOFI}^{(2)}$ in the simulation-based scenario is compatible with $1$, confirming the inadequacy of this approach in low-light scenarios, in particular when the condition in Eq. (\ref{Fanofactor_detected}) is fulfilled. As a result, the super-resolution enhancement of $\text{SOFI}^{(2)}-\text{SIM}$ does not reach the theoretical value of $\sqrt{2}+2$. Nevertheless, both $\text{QSIPS}^{(2)}$ and $\text{QSIPS}^{(2)}-\text{SIM}$ simulations match their theoretical expected resolution factors.

These results demonstrate that QSIPS, including its integration with structured light, offers benefits over standard SOFI.
However, the  Fano factor of the measured photon distribution approaches the threshold value of $2$, in which quantum fluctuation contributions start to become less important. Particularly, in the experiment the brightness was set to achieve a compromise between producing an adequate signal-to-noise ratio for pursuing an accurate quantitative data analysis while also demonstrating QSIPS's clear advantage over SOFI. 
Furthermore, $\text{QSIPS}^{(2)}$ matches the theoretical super-resolution factor of $\sqrt{2}$, while the value for $\text{QSIPS}^{(2)}\text{-SIM}$ is less than the expected one of $2 + \sqrt{2}$.

A possible explanation for this discrepancy is that the theoretical resolution assumes $|\mathbf{p}_{\theta}| = k_{\text{Abbe}}$. However, because the value of $|\mathbf{p}_{\theta}|$ is obtained from an experimental image, working at that frequency is challenging due to the optical system's low transmission at that spatial frequency. In the experiment, we select $|\mathbf{p}_{\theta}| < k_{\text{Abbe}}$, which simplifies SIM reconstruction, but decreases effective resolution enhancement. 

\section{Methods} \label{Methods}
The experimental setup employed in this study is a conventional wide-field fluorescence optical setup with the addition of a Spatial Light Modulator (SLM) for structured illumination.
The system consists of a \SI{532}{\nano\meter} CW laser interacting with the SLM, which imposes a spatially changing phase shift on the incoming light wavefront. The interference pattern is constructed by imposing a 2D cosinusoidal pattern on the SLM, as shown in Eq. (\ref{Intensity_structured}), using a MATLAB script with fixed values of $\theta$ and $\phi$. In this way, the SLM acts as a diffraction gratings producing in the far-field (obtained by a lens) an interference pattern. We select the first two maxima of the interference through a rotating mask,  
which pass through a second focus-adjusting lens and dichroic mirror with a \SI{550}{\nano\meter} cutoff. 
The beam is then focused using a $100\times$ air objective onto a glass substrate containing commercial colloidal CdSe/ZnS QDs, which exhibit a fluorescence emission peak at \SI{620}{\nano\meter} with an FWHM of \SI{20}{\nano\meter}. To create the QDs sample, a toluene solution containing QDs was put on a glass slide, and then diluted with an isopropyl alcohol solution to reduce the QDs concentration. To ensure a uniform distribution of emitters, the sample was spin-coated for \SI{60}{\second} at \SI{100}{rps}. 
The fluorescence is then transmitted via the dichroic mirror, filtered by a \SI{600}{\nano\meter} long-pass and a \SI{700}{\nano\meter} short-pass filter, and captured by a CMOS camera. The Hamamatsu ORCA-QUEST C15550-20UP CMOS camera was utilized in this experiment, characterized by a low readout noise of $0.23$ $\text{e}^-$ and a high quantum efficiency of $85\%$ at \SI{460}{\nano\meter} and $\sim 70\%$ at \SI{620}{\nano\meter}.

\section{Conclusions}\label{Conclusions}

In this paper, we introduced a general quantum approach to super-resolution imaging by photon statistics evaluation, named QSIPS,  which improves by far and extends the classical techniques, such as SOFI.
The QSIPS is applicable to all non-Poissonian emitters and it is optimized at any fluorescence intensity level.   

We established the accuracy of the methods through simulations and experimental work, comparing the performance to those obtained using the standard SOFI technique. Our findings show that QSIPS outperforms the standard SOFI method, particularly in contexts of sub-Poissonian and weak super-Poissonian photon statistics.

Furthermore, the present paper examines the integration of QSIPS with structured light illumination for second-order signals, using both simulation and experiment. Also in this case, the advantages with respect to the equivalent methods are particularly pronounced in low-light conditions. 
Although the QSIPS-SIM approach could not attain the expected theoretical resolution increase due to flaws in frequency vector evaluation, it nonetheless demonstrates novel possibilities for creating a low-power, high-resolution imaging methodology, which might be very useful in future biological applications.\\

\textbf{Funding:} 
This work has received funding by the following projects: Project 20FUN02 POLight of the EMPIR program co-financed by the Participating States and from the European Union’s Horizon 2020 research and innovation program; Qu-Test project, which has received funding from the European Union’s Horizon Europe Research and Innovation Programme under grant agreement
No. 101113901; Project Qutenoise of San Paolo Foundation; Project Phoenicis of "bandi a cascata" PNRR NQSTI; Project INFN QUISS.\\

\textbf{Author contribution:} 
IRB and IPD elaborated the idea and the concept of the experiment that was executed mainly by FP, with contributions from EL, SDC, and PB. The data analysis and simulations were performed by FP.
FP and IPD with the help of IRB elaborated the final theoretical model. MG is the group leader and supervised the project with IRB and IPD. The paper was written with the contribution of all authors, while the first draft was written by FP.\\

\textbf{Disclosures:} The authors declare no conflicts of interest.\\

\textbf{Data availability:} Data underlying the results presented in this paper are not publicly available at this time but may be obtained from the authors upon reasonable request.\\

\textbf{Supplemental document:} See Supplement 1 for supporting content.

\bibliography{Main}

\end{document}


\title{\huge Supplement 1 for "Quantum super-resolution microscopy by photon statistics and structured light"}

\author{F. Picariello}
\email{fabio.picariello@polito.it}
\affiliation{\textit{Politecnico di Torino, Department of Electronics and Telecommunications, Italy}}
\affiliation{\textit{Istituto Nazionale di Ricerca Metrologica, Italy}}

\author{E. Losero}
\affiliation{\textit{Istituto Nazionale di Ricerca Metrologica, Italy}}

\author{S. Ditalia Tchernij} 
\affiliation{\textit{ Physics Dept., University of Torino, Italy}}
\affiliation{\textit{Istituto Nazionale di Fisica Nucleare sez. Torino, Italy}}

\author{P. Boucher}
\affiliation{\textit{Quantonation, Paris, France}}

\author{M.Genovese}
\affiliation{\textit{Istituto Nazionale di Ricerca Metrologica, Italy}}
\affiliation{\textit{ Physics Dept., University of Torino, Italy}}

\author{I. Ruo-Berchera}
\affiliation{\textit{Istituto Nazionale di Ricerca Metrologica, Italy}}

\author{I. P. Degiovanni}
\affiliation{\textit{Istituto Nazionale di Ricerca Metrologica, Italy}}
\affiliation{\textit{ Physics Dept., University of Torino, Italy}}

\begin{abstract}
The supplementary material reports the mathematical proof of the super-resolution model, its intrinsic non-Poissonian characteristics and its relationship with the correlation functions.
\end{abstract}

\maketitle

\section{Mathematical proof}
As reported in the main document, the super-resolved map of order $j$, which is defined as the sum of the $j$-th power of the PSF of each independent emitter, $[\eta_\alpha(\mathbf{r})]^j$, can be evaluated as a linear combination of the cumulants of the overall detected photon distribution from order 1 up to the order $j$:
\begin{equation}
    QSIPS^{(j)}(\mathbf{r}):=\sum_{i=1}^{j}\beta_{i,j}k^{(i)}_\eta (\mathbf{r})=\sum_{\alpha=1}^{N_c}\left[\eta_\alpha(\mathbf{r})\right]^j\left(\sum_{i=1}^{j}\beta_{i,j}z^{(i)}_\alpha\right),\label{QSIPS_relation}
\end{equation}
with:
\begin{equation}
\beta_{i,j} = \frac{1}{j!}\frac{d^j}{dx^j}\prod_{i=0}^{j-1}(x-i)|_{x=0}.
\end{equation}
We refer to such a linear combination of cumulants ($\sum_{i=1}^{j}\beta_{i,j}z^{(i)}_\alpha$) as Sgurzants in the following.

Here we report the mathematical proof for the $\alpha$-th emitter. Therefore, we will demonstrate that:
\begin{equation}
\sum_{i=1}^{j}\beta_{i,j}k_{\alpha,\eta}^{(i)}(\mathbf{r}) = \left[\eta_{\alpha}(\mathbf{r})\right]^j\sum_{i=1}^{j}\beta_{i,j}z_{\alpha}^{(i)},\label{QSIPS_single_emitter}
\end{equation}
where $k_{\alpha,\eta}^{(i)}(\mathbf{r})$ and $z_{\alpha}^{(i)}$ refers to the $i$-th cumulant of the detected and emitted photon distribution of the $\alpha$-th emitter, respectively.
The passage from Eq. (\ref{QSIPS_single_emitter}) to Eq. (\ref{QSIPS_relation}) (i.e. the passage from a single $\alpha$-th emitter to the overall sum of $j$-th power PSF), simply derive from the property of cumulants that ensures, for independent emitters, that the cumulant of the sum is the sum of cumulants, i.e. $k^{(i)}_\eta(\mathbf{r})= \sum_{\alpha=1}^{N_c}  k^{(i)}_{\alpha, \eta}(\mathbf{r})$.  Thus by applying the summation over $\alpha$ to both sides of Eq. (\ref{QSIPS_single_emitter}) we retrieve Eq (\ref{QSIPS_relation}).

To demonstrate Eq. (\ref{QSIPS_single_emitter}), we start considering the cumulant generating function of the emitted and the detected photon distribution for the $\alpha$-th emitter, defined respectively as:
\begin{align}
    \mathcal{Z}_{\alpha}(t)&=\log\left(\sum_{m=0}^{\infty}e^{tm}P_\alpha(m)\right),\label{cumulantemittedalpha}\\
    \mathcal{K}_{\alpha,\eta}(t,\mathbf{r})&=\log\left(\sum_{n=0}^{\infty}e^{tn}\mathcal{P}_\alpha(n,\mathbf{r})\right).\label{cumulantdetectedalpha}
\end{align}
where $\mathcal{P}_\alpha(n,\mathbf{r})$ refers to the probability of detecting $n$ photons emitted form the center $\alpha$ in the position $\mathbf{r}$ and $P_\alpha(m)$ refers to the probability of emitting $m$ photons from the same emitter.

Eq.s (\ref{cumulantemittedalpha}) and (\ref{cumulantdetectedalpha}) can be used to assess the cumulants of the emitted and detected photon distribution for the $\alpha$-th emitter, respectively as:
\begin{align}
    z^{(i)}_{\alpha}&= \frac{d^i\mathcal{Z}_{\alpha}}{dt^i}(t)|_{_{t=0}},\\
    k^{(i)}_{\alpha,\eta}(\mathbf{r})&= \frac{d^i\mathcal{K}_{\alpha,\eta}}{dt^i}(t,\mathbf{r})|_{_{t=0}}.
\end{align}
The detected photon distribution can be written in terms of the emitted photon distribution $P_\alpha(m)$ considered a system of binomial losses as with success probability $\eta_\alpha(\mathbf{r})$:
\begin{equation}
\mathcal{P}_\alpha(n,\mathbf{r})=\sum_{m=n}^{\infty}\binom{m}{n}\left[\eta_\alpha(\mathbf{r})\right]^n\left[1-\eta_\alpha(\mathbf{r})\right]^{(m-n)}P_\alpha(m). \label{Pnalpha}
\end{equation}
Therefore, Eq. (\ref{cumulantdetectedalpha}) can be rewritten using Eq. (\ref{Pnalpha}) as:
\begin{align}
    \mathcal{K}_{\alpha,\eta}(t,\mathbf{r})&=\log\left(\sum_{n=0}^{\infty}e^{tn}\sum_{m=n}^{\infty}\binom{m}{n}\left[\eta_\alpha(\mathbf{r})\right]^n\left[1-\eta_\alpha(\mathbf{r})\right]^{(m-n)}P_\alpha(m)\right) \label{cumulantPm}\\ \nonumber &=\log\left(\sum_{m=0}^{\infty}P_\alpha(m)\sum_{n=0}^{\infty}\binom{m}{n}\left[\eta_\alpha (\mathbf{r})e^t\right]^n\left[1-\eta_\alpha(\mathbf{r})\right]^{(m-n)}\right) \\ \nonumber &= \log \left(\sum_{m=0}^{\infty}\left[1-\eta_{\alpha}(\mathbf{r})+\eta_{\alpha}(\mathbf{r})e^t\right]^mP_\alpha(m)\right),
\end{align}
with $i$ being the cumulant order.

It is useful to consider a different generating function, labeled Sgurzant generating function, $S_{\alpha,\eta}(\gamma_\alpha(t,\mathbf{r}))$, that can be obtained from Eq. (\ref{cumulantemittedalpha}) by making the substitution $e^t\rightarrow \gamma_\alpha(t,\mathbf{r}) = 1-\eta_\alpha(\mathbf{r})+\eta_\alpha(\mathbf{r})e^t$:
\begin{equation}
\mathcal{S}_{\alpha,\eta}(\gamma_\alpha(t,\mathbf{r})) = \log\left(\sum_{m=0}^{\infty}\left[\gamma_\alpha(t,\mathbf{r})\right]^mP_\alpha(m)\right).\label{pica-degio-cumul-gener}
\end{equation} 
Therefore, we can write the definition of $i$-th order detected cumulant for the $\alpha$-th emitter as:
\begin{align}
    k^{(i)}_{\alpha,\eta}(\mathbf{r})= \frac{d^i\mathcal{K}_{\alpha,\eta}}{dt^i}|_{_{t=0}}=\frac{d^i\mathcal{S}_{\alpha,\eta}}{dt^i}|_{_{t=0}}=\frac{d^i\mathcal{S}_{\alpha,\eta}}{dt^i}|_{_{\gamma_\alpha=1}}.\label{cumulantS}
\end{align}
Since we are interested in the $i$-th derivative of the composite function $S_{\alpha,\eta}(\gamma_\alpha(t,\mathbf{r}))$ with respect to the parameter $t$, we can exploit the Faà di Bruno's formula \cite{encinas2003short_suppl}, that states that for a composite function $f(g(x))$ the $i$-th derivative can be written as:
\begin{equation}
    \frac{d^if(g(x))}{dx^i}=\sum_{k=0}^{i}\frac{d^kf(g(x))}{dg^k}B_{i,k}(g^{(1)}(x),g^{(2)}(x),\dots,g^{(i-k+1)}(x)),
\end{equation}
where $B_{i,k}(g^{(1)}(x),g^{(2)}(x),\dots,g^{(i-k+1)}(x))$ denotes partial Bell polynomials \cite{cvijovic2011new_suppl}, defined as:
\begin{equation}
B_{i,k}(g^{(1)}(x),g^{(2)}(x),\dots,g^{(i-k+1)}(x)) = i!\sum_{j_1,j_2,\dots,j_{i-k+1}}\prod_{l=1}^{i-k+1}\frac{\left[g^{(i)}(x)\right]^{j_l}}{(l!)^{j_l}j_l!},
\end{equation}
with $j_1,j_2,\dots,j_{i-k+1}$ non-negative integers that satisfy:
\begin{align}
j_1 + j_2 + \dots j_{i-k+1} &= k,\\
j_1 + 2j_2 + \dots (i-k+1)j_{i-k+1} &= i, \nonumber
\end{align}
and $g^{(1)}(x),g^{(2)}(x),\dots,g^{(i-k+1)}(x)$ represents the derivative of the $g(x)$ function from order $1$ up to order $i-k+1$.

Therefore, employing the Faà di Bruno's formula, Eq. (\ref{cumulantS}) becomes:
\begin{align}
k_{\alpha,\eta}^{(i)}(\mathbf{r}) = \frac{d^i\mathcal{S}_{\alpha,\eta}}{dt^i}|_{_{\gamma_\alpha=1}}= \sum_{k=1}^{i} \frac{d^k\mathcal{S}_{\alpha,\eta}}{d\gamma_\alpha^k}|_{_{\gamma_\alpha=1}}B_{i,k}\left(\gamma_\alpha^{(1)}(0,\mathbf{r}),\dots,\gamma_\alpha^{(i-k+1)}(0,\mathbf{r})\right),\label{diffK}
\end{align}
We find that for any generic order of derivative $i$ of $\gamma_\alpha(t,\mathbf{r})$ respect to the parameter $t$ and evaluated at $t=0$, we obtain:
\begin{equation}
\gamma^{(i)}_\alpha(0,\mathbf{r})=\eta_\alpha(\mathbf{r}),  \hspace{1cm} i \in \mathbb{N}. \label{derivEta}
\end{equation}
Substituting this result into Eq. (\ref{diffK}) leads to:
\begin{equation}
k_{\alpha,\eta}^{(i)}(\mathbf{r}) = \sum_{k=1}^{i} \left[\eta_\alpha(\mathbf{r})\right]^k\frac{d^k\mathcal{S}_{\alpha,\eta}}{d\gamma_\alpha^k}|_{_{\gamma_\alpha=1}}S_{II}(i,k).\label{derivativesecondstirling}
\end{equation}
$S_{II}(i,k)$ represents the Stirling number of the second kind \cite{cvijovic2011new_suppl}:
\begin{equation}
 S_{II}(i,k) = B_{i,k}(1,\dots,1).
\end{equation}
The coefficients $\beta_{i,j}$ coincides with the first type's Stirling number \cite{broder1984r_suppl}:
\begin{equation}
\beta_{i,j} = \frac{1}{j!}\frac{d^j}{dx^j}\prod_{i=0}^{j-1}(x-i)|_{_{x=0}}= S_I(i,j).\label{coeffStirling}
\end{equation}
Using the coefficients $\beta_{i,j}$ from Eq. (\ref{coeffStirling}), we can perform a linear combination up to the order $j$ of the cumulants presented in Eq. (\ref{derivativesecondstirling}):
\begin{align}
\sum_{i=1}^{j}\beta_{i,j}k_{\alpha,\eta}^{(i)}(\mathbf{r})  = \sum_{i=1}^{j} S_I(i,j) \left(\sum_{k=1}^{i}     \left[\eta_\alpha(\mathbf{r})\right]^k\frac{d^k\mathcal{S}_{\alpha,\eta}}{d\gamma_\alpha^k}|_{_{\gamma_\alpha=1}}S_{II}(i,k)\right). \label{almostalmostSR}
\end{align}
Rearranging the sums, Eq. (\ref{almostalmostSR}) becomes:
\begin{align}
\sum_{i=1}^{j}\beta_{i,j}k_{\alpha,\eta}^{(i)}(\mathbf{r}) = \sum_{k=1}^{j}\left[\eta_\alpha(\mathbf{r})\right]^k\frac{d^k\mathcal{S}_{\alpha,\eta}}{d\gamma_\alpha^k}|_{_{\gamma_\alpha=1}}  \left(\sum_{i=k}^{j} S_I(i,j)S_{II}(i,k)\right). \label{almostSR}
\end{align}
Now, using the identity \cite{broder1984r_suppl}:
\begin{equation}
 \sum_{i=k}^{j} S_I(i,j)S_{II}(i,k) = \delta_{j,k},
\end{equation}
we express Eq. (\ref{almostSR}) as:
\begin{align}
\sum_{i=1}^{j}\beta_{i,j}k_{\alpha,\eta}^{(i)}(\mathbf{r}) = \left[\eta_\alpha(\mathbf{r})\right]^j\frac{d^j\mathcal{S}_{\alpha,\eta}}{d\gamma_\alpha^j}|_{_{\gamma_\alpha=1}}.\label{totalSR}
\end{align}
Eq. (\ref{totalSR}) shows that the specific linear combination of cumulants up to order $j$ with coefficients $\beta_{i,j}$ yields terms that are only proportional to the $j$-th power of the PSF, $\left[\eta_\alpha(\mathbf{r})\right]^j$. 

The scaling factor \(d^j \mathcal{S}_{\alpha,\eta} / d \gamma_\alpha^j|_{_{\gamma_\alpha=1}} \) can be written as the same combination of cumulants with coefficients $\beta_{i,j}$, but for the emitted photon distribution instead of the detected one. To demonstrate this, we can pose in Eq. (\ref{cumulantemittedalpha}) $\nu(t) = e^t$, obtaining:
\begin{equation}
    S_\alpha(\nu(t)) = \log\left(\sum_{m=0}^{\infty}\left[\nu(t)\right]^mP_\alpha(m)\right).\label{pica-degio-cumul-gener_rho}
\end{equation}
This generating function coincides with the Sgurzant generating function reported in Eq. (\ref{pica-degio-cumul-gener}). However, in this instance, all the derivatives of $\nu(t)$ are identically equal to $1$ instead of $\eta_\alpha(\mathbf{r})$ as in the detected photon distribution case (see Eq. (\ref{derivEta})):
\begin{equation}
\nu_\alpha^{(i)}(t) = 1, \hspace{1cm} i \in \mathbb{N}.
\end{equation} 
As a result, with the same demonstration steps as done from Eq. (\ref{diffK}) to Eq. (\ref{totalSR}), but relative to the emitted photon distribution $P_\alpha(m)$ instead of the detected one, we obtain:
\begin{equation}
\sum_{i=1}^{j}\beta_{i,j}z_{\alpha}^{(i)} = \frac{d^j\mathcal{S}_{\alpha}}{d\nu^j}|_{_{\nu=1}}=\frac{d^j\mathcal{S}_{\alpha,\eta}}{d\gamma_\alpha^j}|_{_{\gamma_\alpha=1}}.\label{diffcumulantrho}
\end{equation}
Therefore, since $\sum_{i=1}^{j}\beta_{i,j}z_{\alpha}^{(i)}$ are moments of the Sgurzant generating function, we refer to them as Sgurzants.

The last equality of Eq. (\ref{diffcumulantrho}) arises from the identical functional form of Eq.s. (\ref{pica-degio-cumul-gener}) and (\ref{pica-degio-cumul-gener_rho}). Therefore, by combing Eq. (\ref{diffcumulantrho}) with Eq. (\ref{totalSR}), we finally obtain the result in Eq. (\ref{QSIPS_single_emitter}).

In conclusion, the most efficient way to express the QSIPS signal of order $j$ consists in directly writing the Sgurzant generating function of the overall detected photon distribution $\mathcal{P}(N,\mathbf{r})$, given by the convolution of all the emitters at a specific point $\mathbf{r}$ in the detector plane:
\begin{equation}
S_\eta(\rho(t),\mathbf{r}) = \log\left(\sum_{N=0}^{\infty}\left[\rho(t)\right]^N \mathcal{P}(N,\mathbf{r})\right),\label{SfunctionOverall}
\end{equation}
with $\rho(t) = e^t$.

Eq. (\ref{SfunctionOverall}) coincides with the generating function reported in Eq. (\ref{pica-degio-cumul-gener_rho}), albeit for the overall detected photon distribution $\mathcal{P}(N,\mathbf{r})$ rather than the emitted one $P_\alpha(m)$. Therefore, from Eq. (\ref{diffcumulantrho}) we obtain:
\begin{equation}
    QSIPS^{(j)}(\mathbf{r}) = \sum_{i=1}^{j}\beta_{i,j}k_{\eta}^{(i)}(\mathbf{r}) = \frac{d^j\mathcal{S}_{\eta}}{d\rho^j}(\rho,\mathbf{r})|_{_{\rho=1}},
\end{equation}
in agreement with Eq. (\ref{QSIPS_relation}).

\section{Poissonian statistics}

The QSIPS model does not make any assumption on the emitted photon statistics. However, here we demonstrate that considering a Poissonian light source yields to a scaling factor of the super-resolved signal (i.e., the factor multiplying $[\eta_\alpha(\mathbf{r})]^j$ in Eq. (\ref{totalSR})) always zero for every $j>1$.
That means that it is not possible to have super-resolution with Poissonian sources, but only non-Poissonian sources allow to have resolution enhancement.

We can see this by substituting the probability density function $P_\alpha(m)$ of a Poissonian photon source defined as:
\begin{equation}
    P_\alpha(m) = \frac{\left[\lambda_\alpha\right]^m}{m!}e^{-\lambda_\alpha}, \label{PoissonianPM}
\end{equation}
with $\langle m \rangle = \lambda_\alpha$.

Substituting Eq. (\ref{PoissonianPM}) into Eq. (\ref{pica-degio-cumul-gener}) yields to:
\begin{equation}
\mathcal{S}_{\alpha,\eta} = \log\left(\sum_{m=0}^{\infty}\left[\gamma_\alpha\right]^m\frac{\left[\lambda_\alpha\right]^m}{m!}e^{-\lambda}\right).\label{pica_deGiov_poisson}
\end{equation}
Taking the derivative with regard to $\gamma_\alpha(t,\mathbf{r})$ of Eq. (\ref{pica_deGiov_poisson}) yields:
\begin{equation}
\frac{d\mathcal{S}_{\alpha,\eta}}{d\gamma_\alpha} = \frac{1}{\sum\limits_{m=0}^{\infty}\left[\gamma_\alpha\right]^m\frac{\left[\lambda_\alpha\right]^m}{m!}e^{-\lambda_\alpha}}\lambda_\alpha\sum_{m=0}^{\infty}\left[\gamma_\alpha\right]^m\frac{\left[\lambda_\alpha\right]^m}{m!}e^{-\lambda_\alpha}=\lambda_\alpha.
\end{equation}
Since the first derivative is not dependent on the parameter $\gamma_\alpha$, all derivatives with order greater than $1$ of this generating function will be zero, eliminating the super-resolved signal from a Poissonian source, as it is clear from Eq. (\ref{totalSR}).

\section{QSIPS and correlation functions}
In the literature, specifically in \cite{gatto2014beating_suppl}, a mathematical model for super-resolution employing single-photon emitters is proposed and experimentally validated. This model asserts that the super-resolved map of any order \( j \) at a given position \( \mathbf{r} \) in the detector plane can be obtained by multiplying the mean intensity map at the power $j$, \( \langle N(\mathbf{r}) \rangle^j \), with a specific linear combination of correlation functions \cite{titulaer1965correlation_suppl} ($g^{(i)}(\mathbf{r})$, $i \in \mathbb{N}$) up to the \( j \)-th order. The study demonstrates that the first five super-resolved maps, which we will denote as \( SR^{(2)}(\mathbf{r}) \) through \( SR^{(5)}(\mathbf{r}) \), for a system of single-photon emitters, are derived as follows \cite{gatto2014beating_suppl}:
\begin{align}
    SR^{(2)}(\mathbf{r}) &= \langle N(\mathbf{r}) \rangle^2 \left[ 1 - g^{(2)}(\mathbf{r}) \right]= \sum_{\alpha=1}^{N_c}\left[\eta_\alpha(\mathbf{r})\right]^2, \label{SR2} \\
    SR^{(3)}(\mathbf{r}) &= \langle N(\mathbf{r}) \rangle^3 \left[ 1 - \frac{3}{2} g^{(2)}(\mathbf{r}) + \frac{1}{2} g^{(3)}(\mathbf{r}) \right] = \sum_{\alpha=1}^{N_c}\left[\eta_\alpha(\mathbf{r})\right]^3,\label{SR3} \\
    SR^{(4)}(\mathbf{r}) &= \langle N(\mathbf{r}) \rangle^4 \left[ 1 - 2 g^{(2)}(\mathbf{r}) + \frac{1}{2} \left( g^{(2)}(\mathbf{r}) \right)^2 + \frac{2}{3} g^{(3)}(\mathbf{r}) - \frac{1}{3} g^{(4)}(\mathbf{r}) \right] = \sum_{\alpha=1}^{N_c}\left[\eta_\alpha(\mathbf{r})\right]^4,\label{SR4} \\
    SR^{(5)}(\mathbf{r}) &= \langle N(\mathbf{r}) \rangle^5 \left[ 1 - \frac{5}{2} g^{(2)}(\mathbf{r}) + \frac{5}{4} \left( g^{(2)}(\mathbf{r}) \right)^2 + \frac{5}{6} g^{(3)}(\mathbf{r})\right. \label{SR5} \\ \nonumber
    & \quad\quad  - \frac{5}{12} g^{(2)}(\mathbf{r}) g^{(3)}(\mathbf{r}) - \left. \frac{5}{24} g^{(4)}(\mathbf{r}) + \frac{1}{24} g^{(5)}(\mathbf{r}) \right] = \sum_{\alpha=1}^{N_c}\left[\eta_\alpha(\mathbf{r})\right]^5.
\end{align}
where the correlation functions of a generic order $j$ evaluated at zero time delay are defined as:
\begin{equation}
    g^{(j)}(\mathbf{r}) = \frac{\left\langle\prod\limits_{i=0}^{j-1} N(\mathbf{r})(N(\mathbf{r})-i)\right\rangle}{\langle N(\mathbf{r}) \rangle ^k}.
\end{equation}
Here we will show that the first five super-resolved orders of the QSIPS methods can be written in terms of the g functions, providing the same results reported in Eq.s. (\ref{SR2}) - (\ref{SR5}). 

First, we need to consider an expression that allows us to represent cumulants in terms of the raw moments of the overall detected distribution. We can use a recursive formula that connects cumulants to raw moments \cite{smith1995recursive_suppl}, starting with the fact that the first-order cumulant is equivalent to the mean value ($k_\eta^{(1)}(\mathbf{r})=\langle N(\mathbf{r}) \rangle$):
\begin{equation}
    k_\eta^{(j)}(\mathbf{r}) = \left\langle[N(\mathbf{r})]^j\right\rangle - \sum_{i=1}^{j-1} \binom{j-1}{i} k_\eta^{(i)}(\mathbf{r}) \left\langle [N(\mathbf{r})]^{(j-i)} \right\rangle.
\end{equation}
For example, the first five cumulants can be written as:

\begin{align}
    k_\eta^{(1)}(\mathbf{r})&=\langle N(\mathbf{r}) \rangle,\\
    k_\eta^{(2)}(\mathbf{r})&=\left\langle [N(\mathbf{r})]^2 \right\rangle - \langle N(\mathbf{r}) \rangle^2\\
    k_\eta^{(3)}(\mathbf{r})&= \left\langle [N(\mathbf{r})]^3 \right\rangle - 3 \langle N(\mathbf{r}) \rangle \left\langle [N(\mathbf{r})]^2 \right\rangle + 2 \langle N(\mathbf{r}) \rangle^3,\\
    k_\eta^{(4)}(\mathbf{r}) &= \left\langle [N(\mathbf{r})]^4 \right\rangle - 4 \langle N(\mathbf{r}) \rangle \left\langle [N(\mathbf{r})]^3 \right\rangle + 6 \langle N(\mathbf{r}) \rangle^2 \left\langle [N(\mathbf{r})]^2 \right\rangle \nonumber \\
        & \quad  - 3 \langle N(\mathbf{r}) \rangle^4 \nonumber  - 3 \left(\left\langle [N(\mathbf{r})]^2 \right\rangle - \langle N(\mathbf{r}) \rangle^2 \right)^2,\\
    k_\eta^{(5)}(\mathbf{r}) &= \left\langle [N(\mathbf{r})]^5 \right\rangle - 5 \langle N(\mathbf{r}) \rangle \left\langle [N(\mathbf{r})]^4 \right\rangle + 10 \langle N(\mathbf{r}) \rangle^2 \left\langle [N(\mathbf{r})]^3 \right\rangle \nonumber \\
        & \quad - 10 \langle N(\mathbf{r}) \rangle^3 \left\langle [N(\mathbf{r})]^2 \right\rangle + 4 \langle N(\mathbf{r}) \rangle^5 \nonumber \\
        & \quad - 10 \left( \left\langle [N(\mathbf{r})]^3 \right\rangle - 3 \langle N(\mathbf{r}) \rangle \left\langle [N(\mathbf{r})]^2 \right\rangle + 2 \langle N(\mathbf{r}) \rangle^3 \right)\left( \left\langle [N(\mathbf{r})]^2 \right\rangle - \langle N(\mathbf{r}) \rangle^2 \right).
\end{align}
Particularly we have:
\begin{itemize}
    \item \textbf{Second-order:}
    \begin{align}
        QSIPS^{(2)}(\mathbf{r}) &= k^{(2)}_\eta(\mathbf{r}) - k^{(1)}_\eta(\mathbf{r}) \\
        & = \left[\left\langle [N(\mathbf{r})]^2 \right\rangle - \langle N(\mathbf{r}) \rangle^2\right] - \langle N(\mathbf{r}) \rangle \nonumber \\
        & = (-1) \langle N(\mathbf{r}) \rangle^2 \left[1 - \frac{\left\langle [N(\mathbf{r})]^2 \right\rangle - \langle N(\mathbf{r}) \rangle}{\langle N(\mathbf{r}) \rangle^2}\right] \nonumber \\
        & = (-1) \langle N(\mathbf{r}) \rangle^2 \left[1 - g^{(2)}(\mathbf{r})\right]=(-1)\left(SR^{(2)}(\mathbf{r})\right)\nonumber.
    \end{align}
    \item \textbf{Third-order:}
    \begin{align}
        QSIPS^{(3)}(\mathbf{r}) &= k^{(3)}_\eta(\mathbf{r}) - 3k^{(2)}_\eta(\mathbf{r}) + 2k^{(1)}_\eta(\mathbf{r}) \\
        & = \left[\left\langle [N(\mathbf{r})]^3 \right\rangle - 3 \langle N(\mathbf{r}) \rangle \left\langle [N(\mathbf{r})]^2 \right\rangle + 2 \langle N(\mathbf{r}) \rangle^3\right] \nonumber \\
        & \quad - 3 \left[\left\langle [N(\mathbf{r})]^2 \right\rangle - \langle N(\mathbf{r}) \rangle^2\right] + 2 \langle N(\mathbf{r}) \rangle \nonumber \\
        & = 2 \langle N(\mathbf{r}) \rangle^3 \left[1 - \frac{3}{2}\left( \frac{\left\langle [N(\mathbf{r})]^2 \right\rangle - \langle N(\mathbf{r}) \rangle}{\langle N(\mathbf{r}) \rangle^2}\right) \right. \nonumber \\
        & \quad \left.+ \frac{1}{2}\left( \frac{\langle N^3(\mathbf{r}) - 3 N^2(\mathbf{r}) + 2 N(\mathbf{r}) \rangle}{\langle N(\mathbf{r}) \rangle^3}\right)\right] \nonumber \\
        & = 2 \langle N(\mathbf{r}) \rangle^2 \left[1 - \frac{3}{2} g^{(2)}(\mathbf{r}) + \frac{1}{2} g^{(3)}(\mathbf{r}) \right]=2 \left(SR^{(3)}(\mathbf{r})\right)\nonumber.
    \end{align}
    \item \textbf{Fourth-order:}
    \begin{align}
        QSIPS^{(4)}(\mathbf{r}) &= k^{(4)}_\eta(\mathbf{r}) - 6 k^{(3)}_\eta(\mathbf{r}) + 11 k^{(2)}_\eta(\mathbf{r}) - 6 k^{(1)}_\eta(\mathbf{r}) \\
        & = \left[ \left\langle [N(\mathbf{r})]^4 \right\rangle - 4 \langle N(\mathbf{r}) \rangle \left\langle [N(\mathbf{r})]^3 \right\rangle + 6 \langle N(\mathbf{r}) \rangle^2 \left\langle [N(\mathbf{r})]^2 \right\rangle \right. \nonumber \\
        & \quad \left.  - 3 \langle N(\mathbf{r}) \rangle^4 \nonumber  - 3 \left(\left\langle [N(\mathbf{r})]^2 \right\rangle - \langle N(\mathbf{r}) \rangle^2 \right)^2 \right] \nonumber \\
        & \quad - 6 \left[ \left\langle [N(\mathbf{r})]^3 \right\rangle - 3 \langle N(\mathbf{r}) \rangle \left\langle [N(\mathbf{r})]^2 \right\rangle + 2 \langle N(\mathbf{r}) \rangle^3 \right] \nonumber \\
        & \quad + 11 \left[ \left\langle [N(\mathbf{r})]^2 \right\rangle - \langle N(\mathbf{r}) \rangle^2 \right] - 6 \langle N(\mathbf{r}) \rangle \nonumber \\
        & = (-6) \langle N(\mathbf{r}) \rangle^4 \left[ 1 - 2 \left(\frac{\left\langle [N(\mathbf{r})]^2 \right\rangle - \langle N(\mathbf{r}) \rangle}{\langle N(\mathbf{r}) \rangle^2}\right) \right. \nonumber \\
        & \quad + \frac{1}{2} \left( \frac{\left\langle [N(\mathbf{r})]^2 \right\rangle - \langle N(\mathbf{r}) \rangle}{\langle N(\mathbf{r}) \rangle^2} \right)^2 \nonumber \\
        & \quad + \frac{2}{3} \left(\frac{\left\langle [N(\mathbf{r})]^3 \right\rangle - 3 \left\langle [N(\mathbf{r})]^2 \right\rangle + 2 \langle N(\mathbf{r}) \rangle}{\langle N(\mathbf{r}) \rangle^3}\right) \nonumber \\
        & \quad \left.- \frac{1}{6} \left(\frac{\left\langle [N(\mathbf{r})]^4 \right\rangle - 6 \left\langle [N(\mathbf{r})]^3 \right\rangle + 11 \left\langle [N(\mathbf{r})]^2 \right\rangle - 6 \langle N(\mathbf{r}) \rangle}{\langle N(\mathbf{r}) \rangle^4}\right)  \right] \nonumber \\
        & = (-6) \langle N(\mathbf{r}) \rangle^4 \left[ 1 - 2 g^{(2)}(\mathbf{r}) + \frac{1}{2} \left( g^{(2)}(\mathbf{r}) \right)^2  \right. \\ \nonumber &\quad \left. +\frac{2}{3} g^{(3)}(\mathbf{r}) - \frac{1}{6} g^{(4)}(\mathbf{r}) \right]=(-6)\left(SR^{(4)}(\mathbf{r})\right) \nonumber.
    \end{align}
    \item \textbf{Fifth-order:}
    \begin{align}
        QSIPS^{(5)}(\mathbf{r}) &= k^{(5)}_\eta(\mathbf{r}) - 10 k^{(4)}_\eta(\mathbf{r}) + 35 k^{(3)}_\eta(\mathbf{r}) - 50 k^{(2)}_\eta(\mathbf{r}) + 24 k^{(1)}_\eta(\mathbf{r}) \\
        & = \left[ \left\langle [N(\mathbf{r})]^5 \right\rangle - 5 \langle N(\mathbf{r}) \rangle \left\langle [N(\mathbf{r})]^4 \right\rangle \right. + 10 \langle N(\mathbf{r}) \rangle^2 \left\langle [N(\mathbf{r})]^3 \right\rangle \nonumber \\
        & \quad - 10 \langle N(\mathbf{r}) \rangle^3 \left\langle [N(\mathbf{r})]^2 \right\rangle + 4 \langle N(\mathbf{r}) \rangle^5 \nonumber \\
        & \quad \left. - 10 \left( \left\langle [N(\mathbf{r})]^3 \right\rangle - 3 \langle N(\mathbf{r}) \rangle \left\langle [N(\mathbf{r})]^2 \right\rangle + 2 \langle N(\mathbf{r}) \rangle^3 \right)\left( \left\langle [N(\mathbf{r})]^2 \right\rangle - \langle N(\mathbf{r}) \rangle^2 \right)  \right] \nonumber \\
        & \quad - 10 \left[ \left\langle [N(\mathbf{r})]^4 \right\rangle - 4 \langle N(\mathbf{r}) \rangle \left\langle [N(\mathbf{r})]^3 \right\rangle \right. + 6 \langle N(\mathbf{r}) \rangle^2 \left\langle [N(\mathbf{r})]^2 \right\rangle - 3 \langle N(\mathbf{r}) \rangle^4 \nonumber \\
& \quad \left. - 3 \left( \left\langle [N(\mathbf{r})]^2 \right\rangle - \langle N(\mathbf{r}) \rangle^2 \right)^2 \right] + 35 \left[ \left\langle [N(\mathbf{r})]^3 \right\rangle - 3 \langle N(\mathbf{r}) \rangle \left\langle [N(\mathbf{r})]^2 \right\rangle + 2 \langle N(\mathbf{r}) \rangle^3 \right] \nonumber \\
& \quad - 50 \left[ \left\langle [N(\mathbf{r})]^2 \right\rangle - \langle N(\mathbf{r}) \rangle^2 \right] + 24 \langle N(\mathbf{r}) \rangle \nonumber \\
& = 24 \left\langle [N(\mathbf{r})]^5 \right\rangle \left[1 - \frac{5}{2} \left(\frac{\left\langle [N(\mathbf{r})]^2 \right\rangle - \langle N(\mathbf{r}) \rangle}{\left\langle [N(\mathbf{r})]^2 \right\rangle}\right) \right. + \frac{5}{4} \left(\frac{\left\langle [N(\mathbf{r})]^2 \right\rangle - \langle N(\mathbf{r}) \rangle}{\left\langle [N(\mathbf{r})]^2 \right\rangle} \right)^2 \nonumber \\
& \quad + \frac{5}{6} \left(\frac{\langle N^3(\mathbf{r}) - 3 N^2(\mathbf{r}) + 2 N(\mathbf{r}) \rangle}{\left\langle [N(\mathbf{r})]^3 \right\rangle}\right) \nonumber \\
& \quad - \frac{5}{12} \left( \frac{\left\langle [N(\mathbf{r})]^2 \right\rangle - \langle N(\mathbf{r}) \rangle}{\left\langle [N(\mathbf{r})]^2 \right\rangle} \right) \left( \frac{\langle N^3(\mathbf{r}) - 3 N^2(\mathbf{r}) + 2 N(\mathbf{r}) \rangle}{\left\langle [N(\mathbf{r})]^3 \right\rangle} \right) \nonumber \\
& \quad - \frac{5}{24} \left(\frac{\left\langle [N(\mathbf{r})]^4 \right\rangle - 6 \left\langle [N(\mathbf{r})]^3 \right\rangle + 11 \left\langle [N(\mathbf{r})]^2 \right\rangle - 6 \langle N(\mathbf{r}) \rangle}{\left\langle [N(\mathbf{r})]^4 \right\rangle}\right) \nonumber \\
& \quad \left. + \frac{1}{24} \left(\frac{\left\langle [N(\mathbf{r})]^5 \right\rangle - 10 \left\langle [N(\mathbf{r})]^4 \right\rangle + 35 \left\langle [N(\mathbf{r})]^3 \right\rangle - 50 \left\langle [N(\mathbf{r})]^2 \right\rangle + 24 \langle N(\mathbf{r}) \rangle}{\left\langle [N(\mathbf{r})]^5 \right\rangle} \right) \right] \nonumber \\
& = 24 \langle N(\mathbf{r}) \rangle^5 \left[ 1 - \frac{5}{2} g^{(2)}(\mathbf{r}) + \frac{5}{4} \left( g^{(2)}(\mathbf{r}) \right)^2 \right.  + \frac{5}{6} g^{(3)}(\mathbf{r}) - \frac{5}{12} g^{(2)}(\mathbf{r}) g^{(3)}(\mathbf{r}) \nonumber \\
& \quad  \left. - \frac{5}{24} g^{(4)}(\mathbf{r}) \left. \right.  + \frac{1}{24} g^{(5)}(\mathbf{r}) \right]=24\left(SR^{(5)}(\mathbf{r})\right) \nonumber.
\end{align}
\end{itemize}

It can be observed that aside from a constant factor that scales with \((-1)^{(j-1)}(j-1)!\), which does not affect the super-resolution enhancement, the formulations in terms of cumulants and correlation functions coincide. This confirms that the results reported in \cite{gatto2014beating_suppl}, proposed and demonstrated for single-photon emitters, can be more generally expressed using the QSIPS model.

In general, any super-resolution order expressed via QSIPS can be formulated using correlation functions. However, whereas QSIPS utilizes a straightforward linear combination of cumulants, employing correlation functions results in a complex combination of these functions.

\bibliography{supplement}